\documentclass{iopart}

\usepackage{epsfig}

\input{amssym.def}
\newsymbol\lesssim 132E
\newsymbol\gtrsim 1326

\newcommand{\sgn}{{\rm sgn}}
\newcommand{\bt}{\bar \theta}
\newcommand{\bsi}{{\bar \sigma}_+}
\newcommand{\bSp}{{\bar \Sigma}_+}
\newcommand{\ba}{\bar A}
\newcommand{\bk}{\bar K}
\newcommand{\bx}{\bar{x}}

\begin{document}

\jl{6}

\title{Timelike self-similar spherically symmetric perfect-fluid 
models}

\author{Martin Goliath\dag, Ulf S Nilsson\dag \ and 
        Claes Uggla\ddag \\
{\dag \ Department of Physics, Stockholm University, Box 6730,
  S-113 85 Stockholm, Sweden} \\
{\ddag \ Department of Physics, Lule{\aa} University of
  Technology, S-951 87 Lule{\aa}, Sweden}}
 
\begin{abstract}
Einstein's field equations for timelike self-similar spherically symmetric 
perfect-fluid models are investigated. The field equations are rewritten as a 
first-order system of autonomous differential equations. Dimensionless 
variables are chosen in such a way that the number of equations in the 
coupled system is reduced as far as possible and so that the reduced phase 
space becomes compact and regular. The system is subsequently analysed 
qualitatively using the theory of dynamical systems.

Using this approach, we obtain a clear picture of the full phase space and 
the full space of solutions. Solutions of physical interest, e.g. the 
solution associated with criticality in black hole formation, are easily 
singled out. We also discuss the various `band structures' that are 
associated with certain one-parameter sets of solutions.
\end{abstract}
 
\pacs{0420, 0420J, 0440N, 9530S, 9880H}

\maketitle

\section{Introduction}
Spherically symmetric perfect-fluid models admitting a homothetic Killing 
vector has attracted considerable attention during the last couple of decades 
(see, e.g., \cite{carcola,carcolb} and references therein). The main reason 
for this is that the homothetic Killing vector reduces the field equations of 
spherically symmetric perfect-fluid models from partial to ordinary 
differential equations. This simplification does not mean that these 
so-called self-similar models do not contain interesting physical phenomena 
-- quite the contrary. The possibility of violation of the cosmic censorship 
hypothesis \cite{oripir} and the existence of sonic surfaces (shock waves) 
\cite{bogoyavlensky,caryah} are some examples. The growth of primordial black 
holes and the evolution of voids can also be studied using self-similar 
perturbations of the flat Friedmann-Lemaitre-Robertson-Walker (FLRW) model 
\cite{caryah,carrhaw,bicknell1}. In recent years it has become clear that 
self-similar models are of crucial importance for the understanding of black 
hole formation through gravitational collapse. In the full space of 
spherically symmetric solutions, certain self-similar solutions are 
separatrices between dispersive solutions and solutions developing black 
holes \cite{evacol,maison,hara}. 

Self-similar spherically symmetric perfect-fluid models exhibit several 
preferred geometric structures. There are two preferred directions 
corresponding to the 4-velocity of the fluid and the homothetic Killing 
vector, respectively. Adapting the coordinates to the 4-velocity of the fluid 
leads to the `comoving fluid approach' 
\cite{cahtau71,caryah,bicknell1,bicknell2,foglizzo}. Adapting the coordinates 
to the homothetic Killing vector leads to the `homothetic approach' 
\cite{bogoyavlensky,anile}. The spherical symmetry surfaces also constitute a 
preferred geometric structure. The area of the spherical symmetry surfaces is 
used as a coordinate in the `Schwarzschild approach' 
(see, e.g., \cite{oripir}). There are also some other possibilities, e.g. 
synchronous coordinate systems and null coordinates, which are useful for 
discussing singularities \cite{bogoyavlensky}. All these approaches are 
complementary and a full understanding probably relies on a combination of 
results obtained in the various approaches. 

In this paper we will use the diagonal homothetic approach. This has the 
advantage of yielding equations which are very similar to those of 
hypersurface homogeneous models. There exists a wealth of literature on how 
to treat such models. Thus many of the ideas obtained in this area can be 
carried over into the realm of self-similar spherically symmetric models. 
However, there are of course disadvantages as well. The symmetry surfaces 
will in general change causality. Thus, in the diagonal homothetic approach 
the spacetime has to be covered with two coordinate patches; one for when the 
homothetic Killing vector is spacelike and one for when it is timelike. Then 
these two regions must be joined where the homothetic Killing vector is null. 

In a previous paper \cite{sss} we considered the spatially self-similar 
(SSS) spherically symmetric models. In the present paper we consider the 
case of timelike symmetry surfaces, constituting the so-called timelike 
self-similar (TSS) spherically symmetric models. We obtain a full phase space 
picture which leads to a more complete understanding of these models than has 
been previously obtained. Thus the work presented here confirms, corrects and 
expands the analysis of several authors 
\cite{oripir,bogoyavlensky,bicknell1,maison,hara,bicknell2,foglizzo}.

The TSS spherically symmetric models are characterized by a four-dimensional 
homothetic symmetry group $H_4$ acting multiply transitively on 
three-dimensional timelike hypersurfaces. The line element, written in 
diagonal form where one of the coordinates is adapted to the homothetic 
symmetry, takes the form \cite{bogoyavlensky}
\begin{equation}
   d{\tilde s}^2 = e^{2t} ds^2 = e^{2t}\left[
   -D_1^2(x)dt^2 + dx^2 + D_2^2(x)\left( d\theta^2+
   \sin^2{\theta}d\varphi^2 \right)\right] .
\end{equation}
We will consider perfect-fluid models. The energy momentum tensor,
${\tilde T}_{ab}$, is given by
\begin{equation}
  {\tilde T}_{ab} = {\tilde \mu}u_a u_b + {\tilde p}(u_a u_b + g_{ab}) ,
\end{equation}
where $\tilde{\mu}$ is the energy density, $\tilde{p}$ is the pressure, and
$u^a$ the 4-velocity of the fluid. We will assume
\begin{equation}
   \tilde{p}=\left( \gamma-1 \right)\tilde{\mu} ,
\end{equation}
as an equation of state where the parameter $\gamma$ takes values in the 
interval $1 < \gamma < 2$, which include radiation ($\gamma = \case43$). 
We have excluded dust ($\gamma = 1$) and stiff fluids ($\gamma = 2$) as they 
behave quite differently compared to those in the above interval and need 
special treatment. The dust solutions are explicitly known (these models
are just special cases of the general spherically symmetric dust solutions, 
which are all known explicitly, see, e.g., \cite{kramer,carcola}).

The outline of the paper is the following. In section 2 Einstein's field 
equations are rewritten in terms of a dimensionless set of variables, in order 
to obtain a maximal reduction of the coupled system of ordinary differential 
equations. The variables are chosen so that they take values in a compact 
phase space. In section 3 the equations are subsequently analysed by using 
the theory of dynamical systems. In section 4 a numerical investigation is 
undertaken and global dynamical features are considered. Appendix A describes 
the properties of the fluid congruence and also the condition for the 
spacetimes to belong to Petrov type 0. Appendix B gives the relation between 
various coordinates and the relation between the present variables and other 
variables used in the literature.

\section{The dynamical system}
Following \cite{ssslrs}, we introduce
\begin{eqnarray}
  D_1 &=& B_1{}^{-1} = e^{\beta^0 - 2\beta^+} , 
  \qquad D_2 = B_2{}^{-1} = e^{\beta^0 + \beta^+} , \\
  \theta &=& 3{\dot \beta}^0 , \qquad 
  \sigma_+ = 3{\dot \beta}^+ , \nonumber
\end{eqnarray}
where a dot denotes $d/dx$. The quantities $\theta$ and $\sigma_+$ 
describe the kinematical properties of the normal congruence of the symmetry 
surfaces in the static $(M,ds^2)$ spacetime that is conformally related to the 
physical spacetime $(M,d{\tilde s}^2)$ with the homothetic factor $e^{2t}$ 
(see, e.g., \cite{hhss}); $\theta$ is the expansion while $\sigma_+$ describes 
the shear of the normal congruence of the symmetry surfaces in $(M,ds^2)$. 
Expressed in an orthonormal 
frame, the fluid 4-velocity is parametrized by 
$u^a = (1,u,0,0)/\sqrt{1-u^2}$. Einstein's field equations, 
${\tilde G}_{ab} = {\tilde T}_{ab}$, and the equations of motion for the 
fluid, ${\tilde T}^{ab}\!_{;b} = 0$, lead to a set of evolution equations for 
$\theta$, $\sigma_+$, $B_1$, $B_2$ and $u$. In addition, a constraint 
equation, and equations for $\mu_t$ and $\dot{\mu}_t$ are obtained. The 
quantity $\mu_t$, related to $\tilde{\mu}$ by
\begin{equation}
  \mu_t = \frac{1+(\gamma-1)u^2}{1-u^2}\,e^{-2t}\,\tilde{\mu} ,
\end{equation}
is the energy density of the fluid as measured by an observer with a 
4-velocity associated with the homothetic symmetry.

As in the SSS case \cite{sss}, we will now choose to use the $\bt, \bsi$ 
variables, defined by
\begin{eqnarray}
  \bt &=& \frac{1}{\sqrt 3}(2\theta - \sigma_+) , \qquad
  \bsi = \frac{1}{\sqrt 3}(-\theta + 2\sigma_+) , \\
  \theta &=& \frac{1}{\sqrt 3}(2\bt + \bsi) , \qquad
  \sigma_+ = \frac{1}{\sqrt 3}(\bt + 2\bsi) . \nonumber
\end{eqnarray}
This is done in order to simplify the constraint, while still keeping the 
quadratic form of the defining equation for $\mu_t$ 
(equation (\ref{eq:muts2}) below). This leads to:

\paragraph{Evolution equations}
\begin{eqnarray}
  \dot{\bt} &=& -\frac{1}{\sqrt 3}\left[
  \bt^2 + \bsi^2 + \bt\bsi - 3B_1^2 - 
  3\frac{(\gamma - 1)(1 - u^2)}{1+(\gamma-1)u^2}\mu_t \right] , \nonumber \\
  \dot\bsi &=& -\frac{1}{\sqrt 3}\left[\bsi^2 + 2\bt\bsi + 
  \frac{3}{2}\frac{(3\gamma - 2) + (2 - \gamma)u^2}
  {1+(\gamma-1)u^2}\mu_t\right] , \nonumber \\ 
  \dot{B_1} &=&\frac{1}{\sqrt 3} \bsi B_1 , \\
  \dot{B_2} &=& -\frac{1}{\sqrt 3}(\bt + \bsi)B_2 , \nonumber \\
  \dot{u} &=& \frac{1 - u^2}{{\sqrt 3}\gamma[u^2 - (\gamma - 1)]}
  \left\{\gamma\left[2(\gamma - 1)\bt + \gamma\bsi\right]u \right. 
  \nonumber \\ 
  & & \left.+{\sqrt 3}\left[(\gamma - 1)(3\gamma - 2) -
  (2 - \gamma)u^2\right]B_1 \right\} . \nonumber
\end{eqnarray}

\paragraph{Constraint equation}
\begin{equation}\label{eq:cons}
   \gamma u\mu_t - \frac{2}{\sqrt 3}
  \left[ 1+(\gamma-1)u^2 \right]\bsi B_1 = 0 .
\end{equation}

\paragraph{Defining equation for $\mu_t$}
\begin{equation}\label{eq:muts2}
  \mu_t = \frac{1}{3}\left[\frac{1+(\gamma-1)u^2}{u^2 + (\gamma - 1)}\right]
          \left[\bt^2 -\bsi^2 - 3B_1^2 - 3B_2^2 \right] .
\end{equation}

\paragraph{Auxiliary equation}
\begin{eqnarray}
  \dot{\mu}_t&=& \frac{\mu_t}{3\left[1+(\gamma-1)u^2\right]
  \left[u^2-(\gamma-1)\right]}\left\{\sqrt{3}
  \left[(4\bt+\bsi)\gamma-6\bt\right]u^2\gamma \right. \nonumber\\
  & & +6(-7\gamma+3\gamma^2+4)uB_1 -6(-5\gamma+2\gamma^2+4)u^3B_1 \nonumber \\
  & & \left.-\sqrt{3}(\gamma-1)(2\bt+\bsi)u^4\gamma -
  \sqrt{3}\bsi\gamma\right\} . 
\end{eqnarray}
The defining equation for $\mu_t$ shows, assuming non-negative energy 
density, that $\bt$ is a `dominant quantity'. Consequently, in order to 
obtain bounded `reduced' variables, we now introduce $\bt$-normalized 
dimensionless variables
$\bSp$, $\ba$, $\bk$ 
\begin{equation}
  \bSp = \frac{\bsi}{\bt},\qquad \ba = \frac{{\sqrt 3}B_1}{\bt},\qquad
  \bk =  3\left(\frac{B_2}{\bt} \right)^2 .
\end{equation}
The density $\mu_t$ is replaced by the density parameter $\Omega_t$,
which is defined by
\begin{equation}
  \Omega_t=\frac{3\mu_t}{\bt^2} .
\end{equation}
The introduction of a dimensionless independent variable $\eta$
\begin{equation}
  \frac{dx}{d\eta}=\frac{{\sqrt 3}}{\bt} ,
\end{equation}
leads to a decoupling of the $\bt$-equation
\begin{equation}\label{eq:q}
  \bt^\prime=-(1+q)\bt ,  \quad
  q=\bSp(1 + \bSp) - \ba^2 - 
  \frac{(\gamma - 1)(1 - u^2)}{1+(\gamma-1)u^2}\Omega_t ,
\end{equation}
where a prime denotes $d/d\eta$. The remaining coupled evolution equations 
can now be written in dimensionless form:

\paragraph{Evolution equations}
\begin{eqnarray}\label{eq:evoeq}
  \bSp^{\prime} &=& -\bSp\left[ 1 - \bSp^2 + \ba^2 +
  \frac{(\gamma - 1)(1 - u^2)}{1+(\gamma-1)u^2}\Omega_t\right] - \nonumber\\
  & & \frac{1}{2}\frac{(3\gamma - 2) + (2 - \gamma)u^2}
  {1+(\gamma-1)u^2}\Omega_t , \nonumber \\
  \ba^{\prime} &=& \left[1 + 2\bSp + \bSp^2 - \ba^2 -
  \frac{(\gamma - 1)(1 - u^2)}{1+(\gamma-1)u^2}\Omega_t\right]\ba , \\
  \bk^{\prime} &=& 2\left[ \bSp^2 - \ba^2 -
  \frac{(\gamma - 1)(1 - u^2)}{1+(\gamma-1)u^2}\Omega_t\right]\bk,\nonumber\\
  u^{\prime} &=& \frac{1-u^2}{\gamma[u^2 - (\gamma - 1)]} \times \nonumber\\
  & &\left\{\gamma \left[2(\gamma - 1) + \gamma\bSp\right]u + 
  \left[(\gamma - 1)(3\gamma - 2) - (2 - \gamma)u^2\right]\ba\right\} ,
  \nonumber
\end{eqnarray}

\paragraph{Constraint equation}
\begin{equation}\label{eq:tssco}
  G \equiv \gamma u \Omega_t - 2\left(1+(\gamma-1)u^2\right)\bSp \ba = 0 .
\end{equation}

\paragraph{Defining equation for $\Omega_t$}
\begin{equation}\label{eq:omega}
  \Omega_t = \frac{1+(\gamma-1)u^2}{u^2 + (\gamma - 1)}
             \left(1 - \bSp^2 - \ba^2 - \bk\right) .
\end{equation}

\paragraph{Auxiliary equation}
\begin{eqnarray}
  {\Omega_t}^{\prime} &=& \frac{\Omega_t}
  {\left[1+(\gamma-1)u^2\right]\left[u^2-(\gamma-1)\right]}\times\nonumber\\
  & & \left\{\left[(4+\bSp)\gamma-6\right]u^2\gamma
  +2(-7\gamma+3\gamma^2+4)u\ba \right. \nonumber \\
  & & -2(-5\gamma+2\gamma^2+4)u^3\ba -(\gamma-1)(2+\bSp)u^4\gamma
  -\bSp\gamma \nonumber\\ 
  & & \left.+2(1+q)\left[1+(\gamma-1)u^2\right]\left[u^2-(\gamma-1)\right]
  \right\} .  
\end{eqnarray}
Equation (\ref{eq:muts2}) makes it impossible for $\bt$ to change sign
when $\mu_t \neq 0$.  If $\bt > 0$, then $\ba \geq 0$, where we have
included the zero boundary value.  The field equations are invariant
under the transformation 
\begin{equation}\label{eq:disc2} 
  (\bSp,\ba,\bk,u)\rightarrow (\bSp,-\ba,\bk,-u) .
\end{equation} 
Thus it is trivial to obtain the case when $\bt < 0$ corresponding to 
$\ba\leq0$, once the case $\ba\geq0$ has been analysed. We will therefore 
assume that $\bt>0$, $\ba\geq0$. Note that both cases are needed in order to 
obtain a global picture.

The surfaces defined by $u^2 = \gamma-1$ are surfaces of non-extendibility of 
solutions. The only way to cross such a sonic surface
analytically is
through the line where the numerator of the $u'$ equation of (\ref{eq:evoeq})
vanishes. This is the {\it sonic line}. To study the behaviour 
of trajectories around this line, we make the following non-monotonic change 
to a new independent variable $\aleph$,
\begin{equation}\label{eq:aleph}
  \frac{d\aleph}{d\eta}=\frac{1}{(\gamma-1) - u^2} .
\end{equation}

The line element can be obtained when $\bk,\ba$ and $\bt$ have been found
through the relations
\begin{equation}
  D_1 = \sqrt{3}(\bt\ba)^{-1} , \qquad 
  D_2 = \sqrt{3}(\bt^2\bk)^{-1/2} , \qquad 
  x = {\sqrt 3}\int \frac{d{\eta}}{\bt} .
\end{equation}

\section{Dynamical systems analysis of the reduced phase space}
The reduced phase space is determined by ($\bSp,\ba,\bk,u$), related by the
constraint $G = 0$, given by equation (\ref{eq:tssco}). As in \cite{sss}, we 
include the boundary in order to obtain a compact phase space. The boundary
is given by a number of invariant submanifolds: 
$\Omega_t=0$, $\ba=0$, $\bk=0$ and $u=\pm1$. The constraint cannot be solved 
globally everywhere. Instead we will follow \cite{ssslrs,hewitt} and 
solve it locally around the equilibrium points (i.e. we will solve the 
linearized constraint for different variables at different equilibrium 
points). This formulation will enable us to achieve a good understanding of 
the global structure of the dynamics of the reduced phase space. 
For an introduction to dynamical systems analysis see, e.g., 
\cite{dynsyst}, ch 4.

\begin{table}
  \caption{Equilibrium points of the TSS phase space.}
  \begin{indented}
    \item[]
    \begin{tabular}{@{}llllll}
      \hline\hline
      & \multicolumn{4}{c}{Variables} \\ \cline{2-5}
      Notation & $\bSp$ & $\ba$ & $\bk$ & $u$ & $\Omega_t$ \\
      \hline
      $K_+^0$ & 1 & 0 & 0 & 0 & 0 \\
      $K_-^0$ & $-1$ & 0 & 0 & 0 & 0 \\
      $C^0$ & 0 & 0 & 1 & 0 & 0 \\
      $T$ & $-2\frac{\gamma-1}{3\gamma-2}$ & 0 & 
      $\frac{\gamma^2+4(\gamma-1)}{(3\gamma-2)^2}$ & 0 &
      $\frac{4(\gamma-1)}{(3\gamma-2)^2}$ \\
      $\tilde{M}^\pm$ & 0 & 1 & 0 & see text & 0 \\
      $SL$ & & see text & & $-\sqrt{\gamma-1}$ & \\
      $K_+^\pm$ & 1 & 0 & 0 & $\pm1$ & 0 \\
      $K_-^\pm$ & $-1$ & 0 & 0 & $\pm1$ & 0 \\
      $M^+$ & 0 & 1 & 0 & 1 & 0 \\
      $M^-$ & 0 & 1 & 0 & $-1$ & 0 \\
      $\cal{H}^-$ & & $\bSp+1$ & 0 & $-1$ & $-2\bSp\ba$ \\
      $C^\pm$ & 0 & 0 & 1 & $\pm1$ & 0 \\
      \hline\hline
    \end{tabular}
  \end{indented}
\end{table}

\begin{table}
  \caption{Linear analysis of equilibrium points of the TSS phase space.}
  \begin{indented}
    \item[]
    \begin{tabular}{@{}llllll}
      \hline\hline
      & & \multicolumn{4}{c}{Eigenvalues} \\ \cline{3-6}
      Notation &  Elim. & $\lambda_1$ & $\lambda_2$ & $\lambda_3$ & 
      $\lambda_4$ \\
      \hline
      $K_+^0$ & $\ba$ & $\frac{7\gamma-6}{\gamma-1}$ & 2 & 
      $-\frac{3\gamma-2}{\gamma-1}$\\
      $K_-^0$ & $\ba$ & $-\frac{2-\gamma}{\gamma-1}$ & 2 & 
      $\frac{2-\gamma}{\gamma-1}$ \\
      $C^0$ & -- & 1 & 2 & $-1$ & $-2$ \\
      $T$ & $\ba$ & $\frac{2-\gamma}{3\gamma-2}$ & see text & see text \\
      $\tilde{M}^\pm$ & $\bk$ & $-2$ & --"-- & --"-- \\
      $SL$ & $\bk$ & 0 & --"-- & --"-- \\
      $K_+^\pm$ & $\ba$ & 4 & 2 & $-\frac{2(3\gamma-2)}{2-\gamma}$ \\
      $K_-^\pm$ & $\ba$ & 0 & 2 & 2 \\
      $M^+$ & $\ba$ & $-4$ & $-2$ & $-\frac{2(5\gamma-6)}{2-\gamma}$ \\
      $M^-$ & $\ba$ & 0 & $-2$ & $-2$ \\
      $\cal{H}^-$ & $\ba$ & 0 & $-2(1+2\bSp)$ & $-2(1+2\bSp)$ \\
      $C^\pm$ & $\bk$ & $-1$ & 1 & $-\frac{4(\gamma-1)}{2-\gamma}$ \\
      \hline\hline
    \end{tabular}
  \end{indented}
\end{table}

\subsection{Equilibrium points and local analysis}
There are numerous equilibrium points of the TSS dynamical system. In 
table 1 the equilibrium points are presented together with $\Omega_t$, which 
will indicate if a point is on the vacuum boundary $\Omega_t=0$ or not. 
Around each equilibrium point we locally eliminate one of the variables by 
solving the constraint to linear order. In table 2, the variables eliminated 
in the local analysis are listed together with the eigenvalues at each point. 
The equilibrium points  are often closely related. Throughout, they are 
denoted as ${\rm Kernel}_{\sgn (\bSp)}^{\sgn (u)}$. When there is no risk for 
confusion we have suppressed $\sgn{({\bSp})}$ or $\sgn (u)$. The kernel 
indicates the interpretation of the point: $M$ and $C$ refer to Minkowski 
spacetime; $K$ indicates a Kasner solution; F is the flat FLRW solution and T 
refers to a static solution discussed below. Note that 
the self-similar solutions under investigation correspond to orbits in the 
interior of phase space that asymptotically approach the various equilibrium 
points. Below, we will comment on some of the equilibrium points.

\paragraph{The equilibrium point $C^0$.}
This point corresponds to the Minkowski spacetime expressed in spherically 
symmetric coordinates. The constraint surface is 
degenerate, i.e. $\nabla G =(0,0,0,0)$. Hence we keep all four eigenvalues. 

\paragraph{The equilibrium point $T$.}
The eigenvalues $\lambda_2$ and $\lambda_3$ are
\begin{equation}
  \lambda_{2,3} = -\frac{1}{2} \pm i\frac{\sqrt{-\gamma^2+44\gamma-36}}
  {2(3\gamma-2)} .
\end{equation}
Note that these eigenvalues are always complex for
$1<\gamma<2$, and that the real part is always negative.
The point $T$ corresponds to a self-similar static solution
associated with many names: 
Tolman \cite{tolman}, Oppenheimer and Volkoff \cite{oppie}, Klein 
\cite{klein} and Misner and Zapolsky \cite{misnerzapolsky}.
The same solution also leads to an orbit in the interior phase space,
corresponding to a different foliation of the spacetime.
This orbit will be referred to as the {\it static} orbit\footnote{In 
\cite{sss}, it was called the `TOVKMZ' orbit.}, 
and will be discussed in section \ref{sec:globinv} below.

\paragraph{The equilibrium points ${\tilde M}^\pm$.}
For these points, the variable $u$ takes the values
\begin{equation}
  u = \frac{\gamma(\gamma-1)\pm\sqrt{(\gamma-1)\left(\gamma^2(\gamma-1)
  + (3\gamma-2)(2-\gamma)\right)}}{2-\gamma} .
\end{equation}
The expressions for $\lambda_2$ and $\lambda_3$ will not be given since they 
are rather complicated. Instead we briefly comment on the stability of the 
points. At $\gamma=1$ both points coincide, but as $\gamma$ increases the 
points move apart. For $\gamma=\case65$ the point $\tilde{M}^+$ passes 
through the point $M^+$ and leaves the physical part of the TSS phase space. 
It should be noted that the point $\tilde{M}^+$ leaves the TSS phase space 
only to appear in the SSS phase space as one of the points $_\pm\tilde{M}$ 
\cite{sss}. The point $\tilde{M}^-$ exists for all values of $\gamma$ in the 
interval $1<\gamma<2$. It is important to note that $\tilde{M}^-$ always has 
$-\sqrt{\gamma-1}<u<0$, while $\tilde{M}^+$ always has $u>\sqrt{\gamma-1}$. 
This behaviour can be seen in figure 2 (c) - (f). All eigenvalues are 
negative in the interval $1<\gamma\leq\case65$ for ${\tilde M}^+$. Thus this 
point is always an attractive node. For ${\tilde M}^-$ two eigenvalues are 
negative and the third is positive in the interval $1<\gamma<2$. Hence this 
point is a saddle with one outgoing eigendirection. 

\paragraph{The equilibrium points $SL$.}
The flows are opposite to each other on the different sides of the sonic 
surfaces at $u^2=\gamma-1$. Thus it follows that orbits, in general, cannot 
pass through these surfaces continuously. It is only possible to cross a 
sonic surface along the sonic lines $SL$, where the numerator of the $u'$ 
equation in (\ref{eq:evoeq}) vanishes. These lines are defined by
\begin{equation}
  \ba={}^{(-)}_{\ +}\,\frac{\gamma\left(2(\gamma-1)+\gamma\bSp\right)}
  {4(\gamma-1)^{3/2}} , \quad 
  u={}^{(+)}_{\ -}\,\sqrt{\gamma-1} ,
\end{equation}
where the lines of equilibrium points are parametrized by $\bSp$. The signs 
within parentheses correspond to the line for which $u=+\sqrt{\gamma-1}$. 
This line of equilibrium points  is located outside the physical phase space 
for $1<\gamma<2$. Consequently the only relevant sonic line is the one at 
$u=-\sqrt{\gamma-1}$. Using the `shock-adapted' independent variable defined 
in (\ref{eq:aleph}) yields the eigenvalues $\lambda_2$ and $\lambda_3$, which 
are solutions to the equation
\begin{equation}
  4(\gamma-1)\lambda^2 + (2-\gamma)\left(b\lambda + c\right) = 0 ,
\end{equation}
where
\begin{eqnarray}
  b&=&-2\gamma\left(2(\gamma-1)+\gamma\bSp\right) , \quad
  c = (2-\gamma)\left[c_1\bSp^2+c_2\bSp+c_3\right] , \nonumber \\
  c_1 &=& 7\gamma^3 - 38\gamma^2 + 36\gamma - 8 , \quad
  c_2 = 8(\gamma-1)(2\gamma^2-7\gamma+4) , \\
  c_3 &=& -4(2-\gamma)(\gamma-1)^2 . \nonumber
\end{eqnarray}
An analysis of the stability of the line shows that it splits into four 
parts, see figure~1. The boundaries of the physical interval of the line are 
determined by $\ba=0$ and $\bk=0$. The line is a saddle at $\ba=0$, then it 
turns into an attractive node containing the static orbit, discussed in 
section \ref{sec:globinv}. This attractive nodal region is called the S.a.n. 
region. The eigendirection associated to the eigenvalue with the smaller 
absolute value dominates the dynamics close to the sonic line \cite{anile}. 
This direction is called the {\it primary} eigendirection. The other 
direction is referred to as the {\it secondary} eigendirection. Subsequently, 
the eigenvalues become complex and the stability changes to an attractive 
focus. Orbits approaching the sonic line in this region will spiral and hit 
the sonic surface infinitely many times, and are thus unphysical 
\cite{bogoyavlensky}. Finally, the stability changes back to an attractive 
node, now containing the flat FLRW orbit, discussed in section 
\ref{sec:globinv}. This attractive nodal region will be called the F.a.n. 
region. The behaviour along the sonic line is similar for all values of 
$\gamma$ in the interval $1<\gamma<2$. Note that the stability is with 
respect to the new independent variable defined in (\ref{eq:aleph}). In the 
following, if not explicitly pointed out, all figures and discussions of 
phase space flows are with respect to the original independent variable. Note 
that there appears to be an error in the stability analysis of Bogoyavlensky, 
which leads to an erroneous picture of the flow on the phase space boundary 
\cite{bogoyavlensky}. At the boundaries of the focal region, the two non-zero 
eigenvalues are equal. These {\it degenerate nodes} will be denoted S.d.n. 
and F.d.n., respectively. For $\gamma=\case43$, both the static orbit and the 
flat FLRW orbit coincide with the corresponding degenerate nodes on either 
side of the focal region. 

\begin{figure}
  \centerline{ \hbox{\epsfig{figure=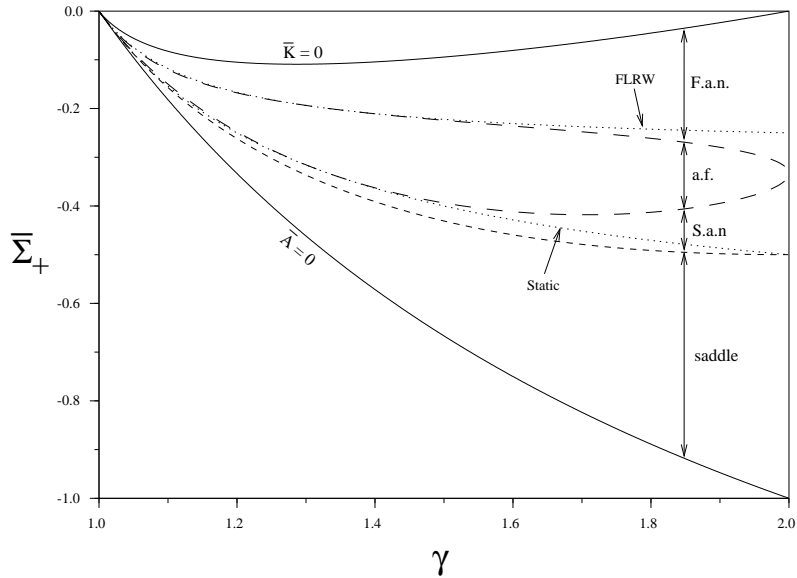, 
    width=0.8\textwidth}}}
  \caption{This figure shows the stability of the sonic line, parametrized 
    by $\bSp$, for $\gamma$ in the interval $1<\gamma<2$. The attractive nodal
    regions are denoted by F.a.n. and S.a.n., while the attractive focus 
    is denoted by a.f. The full curves mark the boundaries of the physical 
    interval for $\bSp$. The dotted curves are the values on the sonic line 
    for the flat FLRW orbit and the static orbit, respectively. A similar 
    figure was presented in \cite{foglizzo}.}
\end{figure}

\paragraph{The equilibrium point $M^+$.}
Note the change in stability of this point at $\gamma=\case65$. This 
bifurcation corresponds to when the point $\tilde{M}^+$ passes through 
$M^+$, leaves the TSS phase space and enters the SSS phase space. 

\paragraph{The equilibrium points $\cal{H}^-$.}
This line of equilibrium points is an artifact of the diagonal homothetic 
approach. It is associated with a causality change of the homothetic 
vector field. To obtain a global picture, one needs to match the TSS phase 
space with the SSS phase space. For the the choice $\bt>0$ ($\bt<0$), 
$\cal{H}^-$ corresponds to the line of equilibrium points $_+\cal{H}^-$ 
($_-\cal{H}^+$) in the SSS phase space \cite{sss}.

\subsection{Invariant submanifolds on the boundary of the TSS phase 
space}\label{sec:invsub}
As stated previously, the boundary is described by a number of invariant 
subsets: $\Omega_t = 0,\ba = 0, \bk = 0,u = \pm 1$. The constraint 
(\ref{eq:tssco}) leads to $u\Omega_t = 0$ when $\ba = 0$. For $u=0$ one 
obtains the reduced equations for the static spherically symmetric 
perfect-fluid models. When $\Omega_t = 0, u\neq0$ one obtains the static 
vacuum equations for a spherically symmetric space time with a test fluid. The 
submanifolds $u=\pm1$ are described by the same equations as the TSS 
spherically symmetric model with a null (e.g. neutrino) fluid. 
The equations for the submanifold $\bk=0$ are the same as the reduced 
equations for the type $_1$I models. The submanifolds are summarized in 
table 3 where we also introduce designations.

\begin{table}
  \caption{The various boundary submanifolds.}
  \begin{indented}
    \item[]
    \begin{tabular}{@{}ll}
      \hline\hline
      Boundary & Restriction \\
      \hline
      $N^\pm$ & $u=\pm1 $\\
      V & $\bSp=0, \Omega=0$\\
      $_1{\rm I}$ & $\bk=0$ \\ 
      $SV_\pm$ & $\ba=0, \Omega=0, \sgn(u) = \pm 1$ \\
      S & $\ba=0, u=0$ \\ 
      \hline\hline
    \end{tabular}
  \end{indented}
\end{table}

We will now describe the dynamical features of the individual boundary 
submanifolds.

\paragraph{The $N^\pm$ submanifolds.}
These submanifolds correspond to Cauchy horizons that are black hole event 
horizons or, in a cosmological context, particle horizons \cite{carcola}. The 
dynamical structure of the $N^-$ submanifold is given in figure 2(a) and the 
corresponding diagram for $N^+$ is given in figure 2(b). The $N^\pm$ 
submanifolds are solvable. The integral describing the various orbits is 
given by
\begin{equation}
  \frac{\bSp \ba}{\left[1-\bSp-\sgn(u)\ba\right]^2} = {\rm constant} .
\end{equation}

\begin{figure}
  \centerline{ \hbox{\epsfig{figure=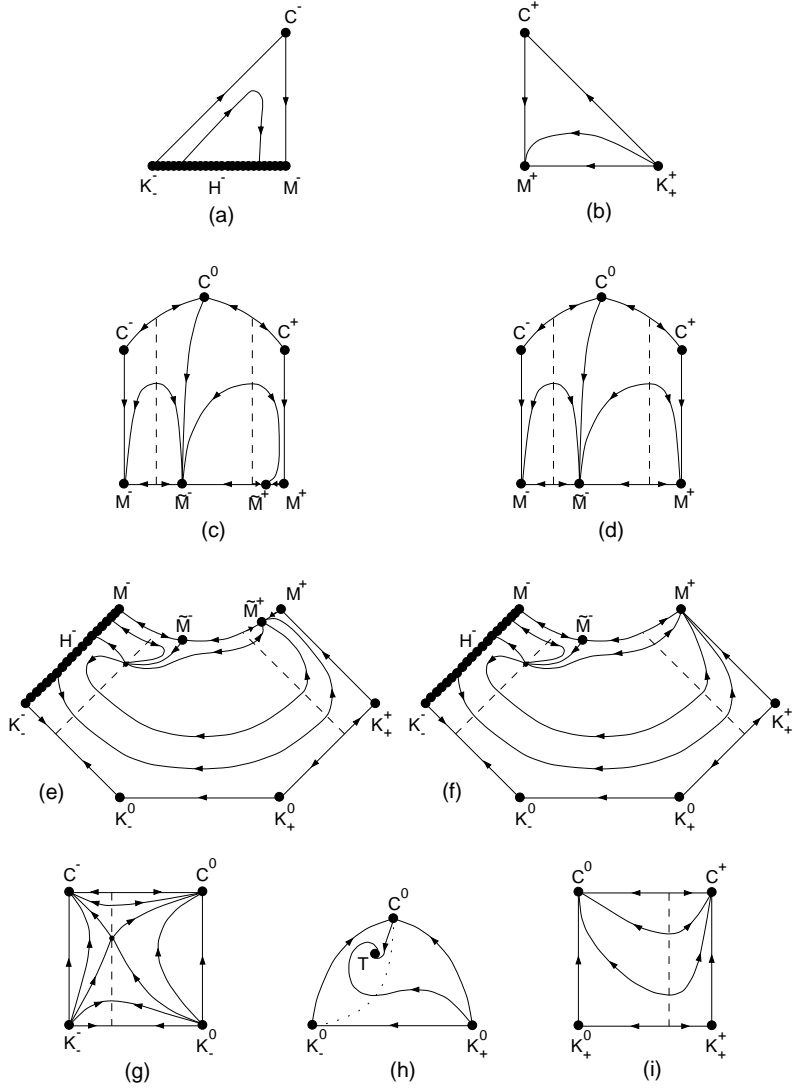, 
    width=0.8\textwidth}}}
  \caption{The phase portraits of the invariant submanifolds that constitute 
    the boundary of the reduced phase space.}
\end{figure}

\paragraph{The vacuum V submanifold.}
The dynamical structure of the V submanifold is given in figure 2(c) 
($\gamma < \case65$) and 2(d) ($\gamma \geq \case65$). Note that $\ba$ is an 
increasing monotonic function, which by the constraint implies that $\bk$ is a 
decreasing monotonic function.

\paragraph{The $_1{\rm I}$ submanifold.}
The dynamical structure of the plane-symmetric type $_1{\rm I}$~submanifold is 
given in figure 2(e) ($\gamma < \case65$) and 2(f) ($\gamma \geq \case65$).

\paragraph{The static vacuum $SV_\pm$ submanifolds.}
The dynamical structure of the $SV_-$~submanifold is given in figure 2(g) 
and the corresponding diagram for $SV_+$ is given in figure 2(i). The 
$SV_\pm$ submanifolds are solvable. The integral describing the various 
orbits is given by
\begin{equation}
  \frac{(1-\bSp)^{3\gamma-2} u^{2(\gamma-1)} (1-u^2)^{2-\gamma}}
  {(1+\bSp)^{2-\gamma} \bSp^{4(\gamma-1)}} = {\rm constant} . 
\end{equation}

\paragraph{The static S submanifold.}
The dynamical structure of the S submanifold is given in figure 
2(h). There exists a monotonic function for this submanifold,
excluding the point $T$. It can be found by using the Hamiltonian methods 
developed in ch 10 of \cite{dynsyst}. It is given by
\begin{equation}
  Z=\frac{\left(3\gamma-2 + 2(\gamma-1)\bSp\right)^2}{\Omega^p\bk^q} , \ 
  p=\frac{4(\gamma-1)^2}{(5\gamma-4)\gamma} , \ 
  q=\frac{(\gamma^2+4(\gamma-1))}{(5\gamma-4)\gamma} , 
\end{equation}
with
\begin{equation}
  \frac{Z'}{Z}=-\frac{2(3\gamma-2)\left[2(\gamma-1) + 
  (3\gamma-2)\bSp\right]^2}
  {\left[3\gamma-2+2(\gamma-1)\bSp\right](5\gamma-4)\gamma} .
\end{equation}
This monotonic function prevents the existence of equilibrium points, periodic 
orbits, recurrent orbits and homoclinic orbits in this region, see e.g. 
\cite{WaHsu}. The static spherically symmetric perfect-fluid models have been
studied qualitatively by Collins using non-compact variables \cite{collins},
but the above monotonic function is, to our knowledge, new. The compact 
formulation taken together with this monotonic function leads to a complete 
understanding of the dynamics of static spherically symmetric models. It is 
worth noting that the orbit from $C^0$ to $T$ corresponds to the only regular 
solution which exists for these models (modulo a scale invariance parameter). 
As will be discussed in section \ref{sec:globinv}, this orbit constitutes 
part of the boundary of a two-dimensional interior submanifold corresponding 
to regular self-similar solutions. The mass function (see \ref{sec:mass}) for 
the S submanifold is
\begin{equation}
  \frac{2m}{R} = 1 - \frac{(1+\bSp)^2}{\bk} .
\end{equation}
The condition $m=0$ defines a curve between $K^0_-$ and $C^0$. This curve is
indicated by the dotted curve in figure 2(h). Both the equilibrium point $T$ 
and the regular solution $C^0$-$T$ are located in the $m\geq 0$ region. The 
corresponding solutions are the only static solutions which have a 
non-negative mass everywhere, since all other orbits come from $K^0_+$, 
located in the $m<0$ region.

\section{Global behaviour}
By appropriately `gluing together' the boundary submanifolds of the previous
section, we obtain the reduced phase space shown in figures 3 and 4. The TSS 
perfect-fluid models correspond to orbits in the interior reduced phase 
space. The $_1{\rm I}$ submanifold makes up the bottom of a `tent' with the 
other submanifolds as `walls'. The point $C^0$ is the top of the `tent'. 
Indicating the stability of the equilibrium points on the boundary, we obtain 
figure 4 for $\gamma > \case65$. For $\gamma \leq \case65$, the point 
${\tilde M}^+$ enters the physical phase space, but is always beyond the 
sonic surface located at $u=\sqrt{\gamma-1}$. It is impossible to 
analytically continue solutions through the sonic surface there, in contrast 
to the one at $u=-\sqrt{\gamma-1}$. Thus the bifurcation at $\gamma=\case65$ 
only affects the `physically uninteresting' part of the phase space, 
$\sqrt{\gamma-1}\leq u \leq 1$. The advantage of a compact and regular phase 
space is now apparent: no parts of the phase space are `crushed'. This is in 
contrast to, e.g., Bogoyavlensky \cite{bogoyavlensky}, where parts of the 
phase space are cut off, while others are located at infinity.
 
The orbits on the boundary corresponding to the eigenvector directions of the
points $\tilde{M}^-$ and $C^0$ are shown in figure 4, as is the static orbit 
along the eigenvector direction of the point $T$ pointing into the phase 
space. The two-dimensional separatrix surface entering the interior phase 
space from $C^0$ and bounded by these orbits will be discussed in section 
\ref{sec:globinv}.

\begin{figure}
  \centerline{\hbox{\epsfig{
    figure=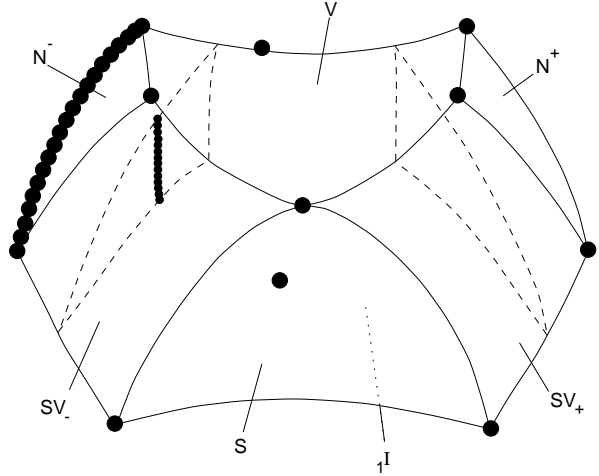,width=0.6\textwidth}}}
  \caption{This picture shows how the boundary of the reduced phase space 
    of the timelike self-similar spherically symmetric models is constructed
    out of the invariant submanifolds discussed in section \ref{sec:invsub}.
    The broken curves describe the sonic surfaces.}
\end{figure}

\begin{figure}
  \centerline{\hbox{\epsfig{figure=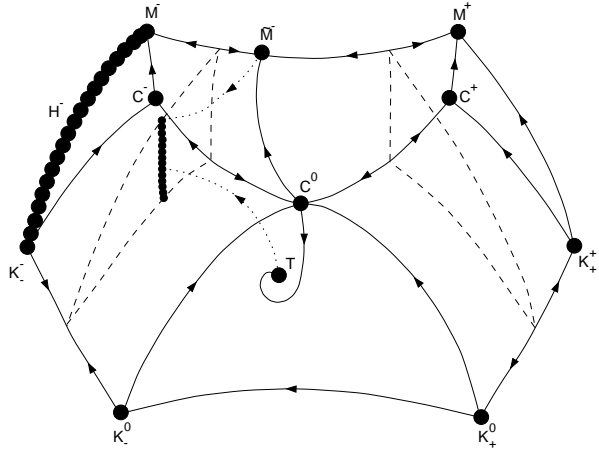, 
    width=0.6\textwidth}}}
  \caption{The global structure of the reduced phase space for 
    $\gamma \geq \case65$. The orbits on the boundary corresponding to the 
    eigenvector directions of the points $\tilde{M}^-$ and $C^0$ are shown, 
    as is the static orbit along the eigenvector direction of the point $T$ 
    pointing into the phase space.}
\end{figure}

\subsection{Monotonic functions}
As indicated above, monotonic functions are important tools for understanding 
the global dynamics. The function \cite{anile}
\begin{equation}
  F = \bar{K}^{2-\gamma}\bSp^{3\gamma-4}\ba^{-\gamma}u^{-2(\gamma-1)}
  \left( 1-u^2 \right)^{-(2-\gamma)} ,
\end{equation}
whose derivative with respect to the independent variable $\eta$ is
\begin{equation}
  F^\prime = \left[\frac{(2-\gamma)(3\gamma-2)(1-u^2)\ba}{\gamma u}\right]F ,
\end{equation}
is easily seen to be monotonic in the regions $u<0$ and $u>0$. This monotonic 
function tells us that all orbits in the $u>0$ part of the phase space come 
from the sonic surface at $u=\sqrt{\gamma-1}$ \cite {bogoyavlensky}. Thus, 
all the `interesting' dynamics takes place for $u<0$.

\subsection{Invariant submanifolds in the interior of the TSS phase 
space}\label{sec:globinv}
A number of trajectories, corresponding to exact solutions of the field 
equations, can be found.

\paragraph{The flat FLRW orbit.}
An orbit corresponding to the flat FLRW solution can be obtained by imposing 
the Petrov type 0 condition, $C=0$ see equation (\ref{eq:weyl}), along with 
vanishing fluid shear. It is described by
\begin{equation}
  \bSp = -\frac{2u^2}{(3\gamma-2)+4u^2} , \quad 
  \ba = -\frac{3 \gamma u}{(3\gamma-2)+4u^2} .
\end{equation}
The flat FLRW orbit passes through the sonic line at an equilibrium point in 
the F.a.n. nodal region. The value of the variable $\bSp$ on the sonic line 
is given by the upper dotted curve in figure 1. The flat FLRW orbit approaches 
the sonic line along an eigendirection. Thus the flat FLRW solution is 
analytic across the sonic surface, as expected (for a discussion about 
differentiability, see the discussion about the ASL submanifold below). The 
corresponding eigendirection is primary for $\gamma<\case43$, and secondary 
for $\gamma>\case43$. This was pointed out in \cite{oripir,foglizzo}.

\paragraph{The static orbit.}
The static orbit, discussed in a previous section, enters the physical phase 
space from the point $T$ and can be found explicitly:
\begin{equation}
  \bSp = -\frac{2(\gamma-1)}{3\gamma-2} , \quad 
  \ba = -\frac{\gamma u}{3\gamma-2} .
\end{equation}
Note that the static solution is not the only solution that is represented 
both as an equilibrium point on the boundary and as a trajectory in the 
interior phase space. This is also the case with the flat FLRW solution in 
the SSS sector, see \cite{sss}. The static orbit passes through the sonic 
line at an equilibrium point in the S.a.n. nodal region. The value of the 
variable $\bSp$ on the sonic line is given by the lower dotted curve in figure 
1. The static orbit approaches the sonic line along an eigendirection. Thus 
the static solution is analytic across the sonic surface, as expected. The 
corresponding eigendirection is secondary for $\gamma<\case43$, and primary 
for $\gamma>\case43$. 

\paragraph{The RC submanifold.}
A requirement on solutions describing collapse is that they should initially 
be regular at the origin. The point $C^0$ is associated with such a regular 
centre \cite{oripir,caryah,foglizzo}. The outgoing eigendirections at this 
point span a surface which enters the reduced phase space. This one-parameter 
family of orbits will be denoted the {\it regular centre} or RC submanifold. 
Its boundary on the V submanifold is the orbit from $C^0$, terminating at 
${\tilde M}^-$ (see figure 2(c) and (d)). The other boundary is the orbit 
from $C^0$ to $T$, continuing into the phase space along the static orbit. 
The orbits in the RC submanifold can be parametrized by the density at the 
centre. For example, Ori and Piran use a parameter $D_0$ (see Appendix B.3). 

The closer an orbit starts to the $C^0-T$ orbit the more it spirals around 
the static orbit. This is to be expected, since the $T$ equilibrium point has 
a pair of complex eigenvalues. The circulation around the static orbit is 
associated with a sign change of the radial 3-velocity of matter $v_R$ 
(see the expression in \ref{sec:schw}). For orbits close to the static orbit, 
$v_R$ changes sign several times. This corresponds to several collapsing and 
expanding spacetime regions enclosing each other. Thus solutions fall into 
different physical families depending on the number of zeroes $n_v$ of $v_R$. 
This defines a band structure of the RC submanifold. Near the V submanifold 
we have `pure collapse' solutions ($n_v=0$). It is obvious that there is an 
infinite number of $n_v$-bands (since $T$ has a pair of complex eigenvalues). 
The flat FLRW solution corresponds to an orbit within the RC submanifold, and 
belongs to the $n_v = 0$ $n_v$-band. The higher $n_v$-band structure 
corresponds to orbits close to the $C^0 - T$ orbit. These orbits and their 
continuation along the static orbit correspond to spacetimes that can be 
described in terms of self-similar coordinates as follows: the spacetime 
(the subsonic part) can be divided into two regions. In the first region the 
solution is approximated by the regular static solution multiplied by the 
homothetic factor. Thus the solution is approximately conformally static in 
this region. In the second region the solution is approximated by a 
perturbation of the static self-similar solution (written in conformally 
static form). This second regime thus corresponds to `quasi-static' 
oscillations around the static solution. The static orbit itself is the only 
orbit starting at $T$, which implies that there are no asymptotically 
quasi-static solutions. However, there is a two-parameter set of solutions 
coming arbitrarily close to the static solution during a limited part of 
their evolution.

\begin{figure}
  \centerline{\hbox{\epsfig{figure=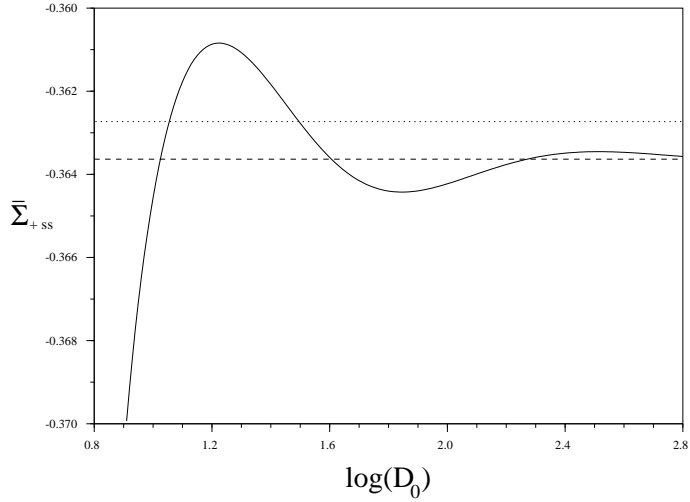, 
    width=0.8\textwidth}}}
  \caption{Values of $\bSp$ at the sonic surface for orbits with 
    $\gamma=1.4$. The broken line is the static value, and the dotted line is 
    the S.d.n. degenerate node. One OP-band starts near the saddle region for 
    $\log(D_0) \approx 0.81$ (not shown) and ends when the curve moves into 
    the focal region at $\log(D_0) \approx 1.05$. This OP-band is the second 
    band, using Ori and Piran's terminology. The third OP-band begins when 
    the curve moves into the S.a.n. nodal region at $\log(D_0) \approx 1.5$. 
    This OP-band is infinitely broad, as the curve never again moves into the 
    focal region.}
\end{figure}

The intersection of the RC submanifold with the sonic surface exhibits another 
band structure in the following sense: when varying the density parameter 
$D_0$ it is found that there are continuous intervals of $D_0$ for which the 
orbits hit the sonic surface at the sonic line. The stability of the sonic 
line determines if an orbit correspond to a physically interesting solution. 
Consequently, `physical' orbits must hit the sonic line in one of the nodal 
regions, or approach the sonic line along the attractive eigendirection of 
the saddle region. These bands were considered by Ori and Piran 
\cite{oripir}, and we will denote them OP-bands. One OP-band consists of 
orbits hitting the F.a.n. nodal region of the sonic line. This is the first 
band in Ori and Piran's terminology. For $1<\gamma\lesssim 1.04$, its upper 
boundary is an orbit approaching the sonic line along a secondary 
eigendirection of the F.a.n. nodal region. The corresponding solution is 
known as the general relativistic Penston-Larson (GRPL) solution 
\cite{oripir}. For $1.04\lesssim\gamma<\case43$, the boundary is the F.d.n. 
degenerate node. This is in accord with the proposition that orbits in the 
neighbourhood of the flat FLRW orbit meet the sonic surface along the sonic 
line for $\gamma<\case43$ \cite{anile}. For higher $\gamma$'s, the boundary 
is the flat FLRW orbit. The first OP-band is separated from higher OP-bands 
by a broad region where the orbits hit the sonic surface at other places than 
the sonic line. Orbits spiraling around the static orbit oscillate around the 
static value when they hit the sonic line. As discussed in connection with 
figure 1, the static orbit coincides with one of the degenerate nodes for 
$\gamma=\case43$. This implies that there will be a large number of OP-bands 
for equations of state with $\gamma$ near $\case43$, because each time 
solutions move into the focal region on the sonic line, an OP-band ends with 
the S.d.n. degenerate node as boundary. When solutions move back into the 
S.a.n. nodal region, the next OP-band begins, again with the S.d.n. 
degenerate node as boundary, see figure 5. As $D_0$ increases, the amplitude 
of the oscillations decreases. Consequently, there is a value $D_{0d}$ such 
that $D_0>D_{0d}$ implies that orbits remain in the static nodal region. 
Thus, there is a finite number of OP-bands, except for $\gamma=\case43$ for 
which the static orbit coincides with the S.d.n. degenerate node. Note that 
the OP-band structure also is affected by whether the orbits hit the sonic 
surface on the sonic line or not. This is the dominant effect for creating 
OP-band structure for soft equations of state. For $1<\gamma\lesssim 1.41$, 
the lower boundary of the second OP-band corresponds to an attractive 
eigendirection in the saddle region. For $1.41\lesssim\gamma\lesssim 1.89$ 
the boundary is a secondary eigendirection in the S.a.n. nodal region, while 
for $\gamma\gtrsim 1.89$ it is the S.d.n. degenerate node (see also the 
discussion about critical behaviour below). The upper boundary changes 
character, depending on the value of $\gamma$. For $1<\gamma\lesssim 1.11$, 
it is a secondary eigendirection in the S.a.n. nodal region. For 
$1.11\lesssim\gamma\lesssim 1.45$, it is the S.d.n. degenerate node. The 
orbits in the second OP-band belong to $n_v$-band $n_v=1$ or $n_v=2$. For 
$\gamma \gtrsim 1.45$, the orbits never move into the focal region on the 
sonic line. The OP-band structure then degenerates to only two bands; the 
first OP-band, containing the flat FLRW solution, and a second band, 
infinitely broad in $D_0$. 

\paragraph{The ASL submanifold.}
In order for the solution to be analytic (or $C^\infty$) at the sonic point, 
the corresponding orbit has to approach the sonic line along an eigendirection 
in one of the nodal regions \cite{bicknell1,bicknell2}, or possibly along the 
attractive eigendirection in the saddle region. The corresponding orbits form 
a submanifold which we will call the ASL submanifold. 

\paragraph{The RCASL submanifold.}
An analytic solution is regular at the centre and analytic at the sonic 
point. These solutions correspond to orbits that belong to the intersection 
of the RC and ASL submanifolds. This intersection forms a submanifold which we 
will denote as the RCASL submanifold. The RCASL submanifold turns out to be a 
discrete set of orbits. Typically, there is only one such solution for each 
$n_v$-band. A convenient numerical way to find orbits belonging to 
$RCASL =RC \cap ASL$ is to start along eigendirections at the sonic line. 
This gives a set of orbits in ASL. For an orbit belonging to RC, $v_R$ should 
tend to zero, as $v_R=0$ at $C^0$. It is thus possible to `fork in' the 
interesting solutions.

Foglizzo and Henriksen \cite{foglizzo} use a set of variables particularly
adapted to numerical studies of orbits between the sonic surface and the 
regular centre. However, we point out that their variables are  non-compact 
and distort the global picture.

\paragraph{Criticality and self-similarity.}
Maison \cite{maison} and Hara {\it et al} \cite{hara,hara2} consider the 
critical behaviour of spherically symmetric perfect-fluid collapse. A 
particular self-similar solution turns out to be of fundamental interest in 
these studies. This solution corresponds to an analytic orbit in the $n_v=1$ 
$n_v$-band, and constitutes the lower boundary with respect to $D_0$ of the 
second OP-band for $1\leq\gamma\lesssim 1.89$. For $\gamma\lesssim 1.41$ it 
approaches the sonic line in the saddle region along the attractive 
eigendirection. In the interval $1.41\lesssim\gamma\lesssim 1.89$, it 
approaches the sonic line along a secondary eigendirection in the S.a.n. 
nodal region. For $\gamma\approx 1.61$, it passes through the point for which 
the static orbit reaches the sonic line. Above this value of $\gamma$, the 
zero in $v_R$ occurs on the opposite side of the sonic point. For 
$\gamma\gtrsim 1.89$, the solution ends in the focal region, and thus becomes 
unphysical, see figure 6. The S.d.n. degenerate node is then the lower 
boundary of the second OP-band.

\begin{figure}
  \centerline{\hbox{\epsfig{figure=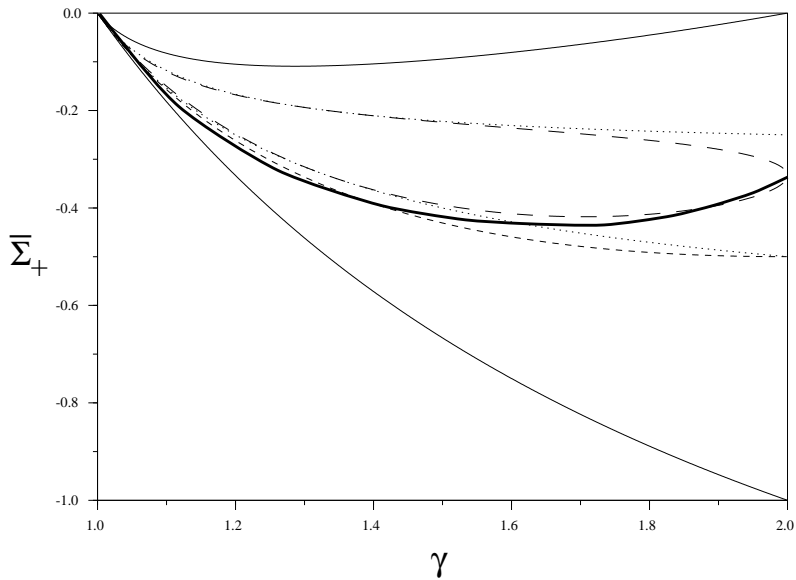, 
    width=0.8\textwidth}}}
  \caption{This figure shows the stability of the sonic line, parametrized 
    by $\bSp$, for $\gamma$ in the interval $1<\gamma<2$. The heavy full curve 
    corresponds to the intersection of the critical solution with the sonic 
    line. For other designations, see figure 1.}
\end{figure}

\paragraph{The F submanifold.}
For a global picture of self-similar spherically symmetric perfect fluid 
models, a careful study of the matching between the TSS and the SSS regions 
is necessary. In our previous work on the SSS region of the phase space 
\cite{sss}, we studied a separatrix surface called the F submanifold. Orbits 
in this submanifold originate from an equilibrium point $_+F$ in the SSS 
region of the phase space. This point corresponds to the flat FLRW solution 
and it gives the submanifold its name. However, the flat FLRW solution is not 
only represented as an equilibrium point on the boundary, but also as an 
orbit within the F submanifold (corresponding to a different foliation of the 
spacetime). The remaining orbits within the F submanifold correspond to 
models that are of considerable physical interest since they can be 
interpreted as density perturbations of the flat FLRW model. The F 
submanifold in the SSS region is divided into two parts; one corresponding to 
orbits lying entirely in the SSS region and one corresponding to orbits 
reaching the ${\cal H}^-$ line of equilibrium points (this equilibrium line 
is denoted $_+{\cal H}^-$ in the SSS region). This latter part of the F 
submanifold corresponds to solutions that can be analytically extended into 
the TSS region.

In order to connect the SSS and TSS phase spaces we use the variables of 
Foglizzo and Henriksen \cite{foglizzo} (these variables arise in a comoving 
context and can therefore be used to deal with the situation where the 
self-similar symmetry surface becomes null, see \ref{sec:vartransf}). Orbits 
corresponding to overdense solutions diverge from the flat FLRW orbit. 
Sufficiently overdense solutions re-enter the SSS region through ${\cal H}^-$ 
and correspond to black hole solutions. However, there is a set of overdense 
orbits reaching the sonic surface. 

Investigating the extendibility of orbits through the sonic surface, we find 
a band structure similar to the OP-band structure of the RC submanifold. We 
refer to these bands as {\it F-bands}. All underdense orbits, from the orbit 
in the $_1{\rm I}$ submanifold to the orbit corresponding to the flat FLRW 
solution, pass the sonic surface through the sonic line. For 
$\gamma\geq\case43$ the limit of this band is the flat FLRW solution. For 
$\gamma<\case43$ there is a set of orbits corresponding to overdense 
solutions in this band. Between the first F-band and the next F-band there is 
a one-parameter subset of orbits that are inextendible and end at the sonic 
surface. This is similar to the region between the first and second OP-band 
of the RC submanifold. Between the subset of inextendible orbits and the 
orbits that return to the SSS region, there is a small band of orbits 
corresponding to overdense solutions that extend through the sonic surface 
via the sonic line. Our numerical investigation indicates that there is just 
one overdense F-band, rather than a hierarchy of several overdense bands (as 
was the case for the overdense OP-bands of RC for $\gamma\lesssim1.45$). The 
interval of the sonic line corresponding to the overdense F-band is wider 
than the corresponding interval covered by these OP-bands. Consequently, 
there are overdense subsonic ($u>-\sqrt{\gamma-1}$) solutions which are 
regular at the sonic surface, but cannot be connected with $C^0$. Numerical 
analysis indicates that these orbits do not recollapse, but have a second 
irregular sonic point. Therefore, they are unphysical \cite{foglizzo}. This 
is in accord with the conjecture of Carr and Coley \cite{carcolb} that 
asymptotically Friedmann solutions which contain black holes are supersonic 
everywhere.

\paragraph{Mass, energy and density fluctuations.}
The mass function $2m/R$ (see equation (\ref{eq:mass}) in \ref{sec:mass}) is 
a physically important quantity. By imposing the condition $m\geq0$, only 
certain parts of the phase space are physical. In figure 2(h), the dotted 
line indicates the condition $m=0$ in the static S submanifold (see the 
discussion of this submanifold in section \ref{sec:invsub}). 

Another interesting quantity is the asymptotic energy per unit mass $E$ of a 
spherical shell (see equation (\ref{eq:ene}) in \ref{sec:mass}). Carr and 
Coley \cite{carcola,carcolb} use this parameter in a general classification 
of self-similar spherically symmetric perfect fluids. We note that $E=0$ for 
the flat FLRW solution, whereas the static solution has $E<0$.

Self-similar density perturbations of the flat FLRW solution can be studied 
covariantly following Ellis and Bruni \cite{ellbru}. The fractional density 
gradient is characterized by a quantity $L$ (see \ref{sec:dens}). For the 
flat FLRW solution, $L$ is zero, as expected. Orbits between the flat FLRW 
orbit and the V submanifold have $L>0$ and correspond to under-dense 
solutions, while orbits on the other side of the flat FLRW orbit have $L<0$, 
corresponding to overdense solutions.

\section{Discussion}
In this paper we have studied timelike self-similar spherically symmetric
perfect-fluid models using a dynamical systems approach. It continues the work
initiated in a previous paper, dealing with the spatially self-similar 
models. By suitably adapting the variables to the homothetic Killing vector, 
the reduced phase space for these models becomes compact and regular, a 
desirable property that has never been obtained before. An advantage of a 
compact and regular phase space is that it enables a more thorough global 
picture of the space of solutions. Also, the current local treatment of the 
constraint makes it possible to avoid `crushing' parts of the phase space. 
The solution space of the TSS models has a much more complicated structure 
than in the SSS case. In particular, the existence of a sonic line in one of 
the sonic surfaces has profound implications for the global structure. The 
sonic surfaces and the sonic line act as `filters' determining 
differentiability of various solutions.

To obtain a full global picture of self-similar spherically symmetric
perfect-fluid models, it is necessary to match TSS and SSS regions. This has 
only been partially discussed here. A more complete discussion will be given 
in a subsequent paper. This will enable us to obtain a more detailed 
understanding of physical phenomena, such as formation of black holes and 
naked singularities.

By allowing an additional coordinate dependence in the metric coefficients, 
the approach of this paper can easily be generalized to non-self-similar 
spherically symmetric models. This is of interest when studying critical 
behaviour in spherically symmetric gravitational collapse. However, the field 
equations will now be partial differential equations, and the phase space 
will no longer be compact. Nevertheless, for small perturbations, this 
generalization of the present approach may be useful.

\section*{Acknowledgments}
We wish to thank B J Carr and A A Coley for interesting discussions and for 
kindly providing us with access to unpublished manuscripts. CU was supported 
by the Swedish Natural Research Council.

\appendix

\section{Interesting physical quantities}

\subsection{Fluid properties and Petrov type conditions}
The fluid expansion $\tilde{\theta}$, the fluid shear $\tilde{\sigma}$, the 
fluid acceleration scalar $\tilde{a}$ and the Weyl scalar $C$ are given by
\begin{eqnarray}
  \tilde{\theta} &=&\frac{e^{-t}}{\sqrt{3}(1-u^2)^{3/2}}\left\{ u^{\prime}
  +(1-u^2)\left[\ba + (2+\bSp)u\right]\right\}\bt , \\
  \tilde{\sigma} &=& \frac{e^{-t}}{3(1-u^2)^{3/2}}\left|
  u^{\prime}-(1-u^2)(1+2\bSp)u\right|\bt , \\
  \tilde{a} &=&\frac{e^{-t}}{3\sqrt{3}(1-u^2)^{3/2}}
  \left| uu^{\prime}+3(1-u^2)(u\ba-\bSp)\right|\bt , \\
  C&=&\sqrt{C_{abcd}C^{abcd}} = \frac{2e^{-2t}}{3\sqrt{3}}
  \left|2\bSp'-(1+q+\bSp)(1+2\bSp)+\bk\right|\bt^2 , \label{eq:weyl} 
\end{eqnarray}
where $q$ was defined in (\ref{eq:q}). Note that the magnetic part 
of the Weyl tensor is identically zero for all models. The spacetime is of 
Petrov type D if $C\neq0$, and of type 0 if $C=0$.

\subsection{Mass and energy}\label{sec:mass}
For a general spherically symmetric spacetime the total mass-energy $m$ 
between the centre distribution and some 2-space of symmetry is defined as
(see, e.g., Misner and Sharp \cite{misnersharp}, Hernandez and Misner 
\cite{hernandez} and Cahill and McVittie \cite{cahill})
\begin{equation}\label{eq:mass}
  \frac{2m}{R} = 1-{\tilde g}^{ab}\frac{\partial R}{\partial x^a}
  \frac{\partial R}{\partial x^b} = \frac{\bk - \left( 1+\bSp\right)^2 + \ba^2}
  {\bk} ,
\end{equation}
where $R=e^t \, D_2(x)$ is the invariant radius of the symmetry surfaces. By 
definition, the matter is outside the gravitational radius $2m$ whenever 
$\frac{2m}{R}<1$ and inside the gravitational radius when $\frac{2m}{R}>1$. 
The latter case is associated with the existence of a black-hole apparent 
horizon or a cosmological apparent horizon at $R=2m$.

It is possible to define the asymptotic energy per unit mass $E$ of a 
spherical shell as \cite{misnersharp}
\begin{equation}\label{eq:ene}
  E = \frac{1}{2}\left[U^2 - \frac{2m}{R}\right] 
  = \frac{1}{2}\left[\frac{(1+\bSp+\ba u)^2}{\bk(1-u^2)} -1\right] ,
\end{equation}
where 
\begin{equation}
  U= \frac{1}{\sqrt{1-u^2}}
  \left(e^{-t}D_1^{-1} \frac{\partial}{\partial t}R + 
      e^{-t} u       \frac{\partial}{\partial x}R \right) 
  = \frac{\ba + u(1+\bSp)}{\sqrt{\bk}\sqrt{1-u^2}} .
\end{equation}

\subsection{Density perturbations}\label{sec:dens}
A covariant approach to density perturbations has been given in 
\cite{ellbru}. The fractional density gradient is defined as
\begin{equation}
  \Delta_a = \left({\Delta{\tilde \mu} \over {\tilde \mu}}\right)_a =
  {\tilde\mu}^{-1} {\tilde h}_a\!^{b} 
  {\tilde \partial}_b{\tilde \mu} .
\end{equation}
In the reduced phase space variables, this leads to 
\begin{equation}
  \Delta_a = \bt\,e^{-t} \, {L \over 1-u^2}\left(-u,1,0,0\right) ,
\end{equation}
where
\begin{equation}
  L = -\frac{2 \ba u\left[u^2+(\gamma-1)\right] +
  \gamma \left[2 u^2 +(1+u^2)\bSp \right]}
  {\sqrt{3} \left[u^2 -(\gamma-1)\right]} .
\end{equation}

\section{Coordinate and variable transformations}\label{sec:vartransf}
\setcounter{equation}{0}
Here we will give transformations to other coordinates and variables which 
have been used in the literature to study self-similar spherically symmetric 
models. The present variables lead to a compact and regular description
everywhere in the TSS sector. This is not the case with previously used 
variables, as is easily seen in the equations below. The simple algebraic 
relations make it easy to identify the points, or even manifolds, where 
breakdowns occur.

\subsection{The variables used in the SSS region}
In a previous paper \cite{sss}, the form of the $\mu_n$ equation led us to 
compactify the variables with respect to $Y=\sqrt{\bt^2 +3 B_2^2}$ rather 
than $\bt$. The resulting set of variables 
$({\bar Q}_0, {\bar Q}_+, {\bar C}_1, v)$ are related to the variables used 
in this paper as follows:
\begin{equation}
  {\bar Q}_0 = \frac{\sgn(\bt)}{\sqrt{1+\bk}} , \
  {\bar Q}_+ = \frac{\bSp \, \sgn(\bt)}{\sqrt{1+\bk}} , \quad
  {\bar C}_1 = \frac{\ba \, \sgn(\bt)}{\sqrt{1+\bk}} , \quad
  v = u^{-1} .
\end{equation}
Note that since $\bt$ is a dominant quantity in the TSS region, $Y$ is even 
more dominant. Thus we could have compactified the variables in the TSS 
region in the same way as in the SSS region with the additional replacement 
$u=1/v$. However, the present choice of compactification result in simpler 
equations. The TSS variables are of course not compact in the SSS region, as 
$\bt$ is not a dominant quantity there. This is easily seen in the above 
relations.

\subsection{Bogoyavlensky's variables}
Bogoyavlensky uses the homothetic approach, solves the constraint globally,
and introduces the variables $Q, w$ and $u$ (as a starting point)
\cite{bogoyavlensky}. This results in an undesirable `crushing' 
of the phase space. Bogoyavlensky's variables are related to the present ones 
by
\begin{equation}
  Q=\frac{\ba}{1+\bSp} , \quad 
  w=-\frac{2\bSp}{1+\bSp} .
\end{equation}

\subsection{The comoving (fluid) approach}
As the fluid 4-velocity singles out a preferred timelike direction in the 
spacetime, it should be beneficial to adapt the coordinates to this direction. 
The line element can be written as
\begin{equation}\label{comoving}
  ds^2 = -e^{\Psi(\lambda)}dT^2 + e^{\Lambda(\lambda)}dX^2 +
  Y(\lambda)^2 T^2d\Omega^2 ,
\end{equation}
where $\lambda=T/X$ and $d\Omega^2 = d\theta^2+\sin^2{\theta}d\varphi^2$.
By defining $T = \exp[t-F(\bx)]$, $X=\exp[t-F(\bx)+\bx]$, where the
function $F(\bx)$ satisfies 
\begin{equation}
  \frac{dF}{d\bx} = -\frac{e^{\Lambda+2\bx}}{e^\Psi-e^{\Lambda+2\bx}} ,
\end{equation}
the line element equation (\ref{comoving}) can be written as
\begin{equation}
  ds^2 = e^{2t}\left( -D_1^2dt^2 + N^2d\bx^2 + D_2^2d\Omega^2\right)
\end{equation}
with
\begin{equation}
  D_1^2 = e^{-2F}\left( e^\Psi - e^{\Lambda+2\bx}\right) , \
  N^2 = \frac{e^{\Lambda+2\bx+\Psi-2F}}{e^\Psi - e^{\Lambda+2x}} , \
  D_2^2 = e^{-2F}Y^2 .
\end{equation}
Using that the fluid velocity is given by $u^2=e^{\Lambda-\Psi + 2\bx}$, the 
above formulas can be written as
\begin{equation}
  \frac{dF}{d\bx} = - \frac{u^2}{1-u^2} , \quad
  D_1^2 = e^{\Psi-2F}\left( 1-u^2 \right) , \quad
  N^2 = \frac{u^2e^{\Psi-2F}}{1-u^2} ,
\end{equation}
with $D_2$ the same as previously. The comoving approach has been used by 
many authors. The most useful variables produced so far to study self-similar 
spherically symmetric collapse of a perfect fluid in this approach are 
probably the ones obtained by Foglizzo and Henriksen \cite{foglizzo} (see also 
Bicknell and Henriksen \cite{bicknell1,bicknell2}). Their variables 
$\left\{ N_{F-H}, \mu_{F-H}, v_{F-H}\right\}$ are related to the present 
ones by
\begin{equation}
  N_{F-H}=\frac{3\gamma\bSp}{2u\ba} , \quad
  \mu_{F-H}=\frac{3\gamma u\left(\bk-\left(1+\bSp\right)^2+\ba^2\right)}
  {2(1-u^2)\bSp\ba} , 
\end{equation}
and $v_{F-H}=u$. Here, we eliminated $\mu_t$ using the constraint equation 
(\ref{eq:cons}). Note the close connection between $\mu_{F-H}$ and  the mass 
function (\ref{eq:mass}). Again it is obvious that these variables are not 
compact. However, they are useful for connecting the SSS and TSS regions
and they are also very well suited for studying the phase space in the 
vicinity of the sonic line.

\subsection{The Schwarzschild approach}\label{sec:schw}
One often represents a spherically symmetric line-element with
its `Schwarzschild' form:
\begin{equation}
\label{schwarz}
  ds^2 = -FdT^2 + GdR^2 + R^2d\Omega^2 ,
\end{equation}
where, in the self-similar case, $F$ and $G$ are functions of $R/T$ 
only. By introducing $R=e^tB_2^{-1}$, $T=e^te^\phi$, where the function 
$\phi$ satisfies the differential equation
\begin{equation}
  \frac{d\phi}{dx}=\frac{\ba^2\bt}{\sqrt{3}\left(1+\bSp\right)} ,
\end{equation}
the line element (\ref{schwarz}) transforms into the diagonal homothetic form.
The metric functions $F$ and $G$ become
\begin{equation}
  F=\frac{3(1+\bSp)^2e^{-2\phi}\bt^{-2}}
  {\ba^2\left[\left(1+\bSp\right)^2-\ba^2\right]} , \quad
  G=\frac{\bk}{\left(1+\bSp\right)^2-\ba^2} .
\end{equation}
Defining $G^{-1}=1-2m/R$ recovers the definition of the mass 
function (\ref{eq:mass}). The radial 3-velocity of matter $v_R$, 
i.e. the speed of the fluid with respect to the surfaces $R=\rm{constant}$, is 
given by
\begin{equation}
  v_R=\frac{\ba+u\left(1+\bSp\right)}{1+\bSp+u\ba} .
\end{equation}
The Schwarzschild approach has been used by several authors. The variables of
Ori and Piran \cite{oripir} are given by
\begin{eqnarray}
  {\cal M}&=&\frac{\bk-\left(1+\bSp\right)^2+\ba^2}{2\bk} , \quad
  D=\frac{\left(1-u^2\right)\bSp\ba}{\gamma\bk u} , \\
  u^r&=&\frac{\ba+u\left(1+\bSp\right)}{\sqrt{\left(1-u^2\right)\bk}} 
  . \nonumber
\end{eqnarray}
The orbits in the RC submanifold are parametrized by 
\begin{equation}
  D_0 \equiv \lim_{R/T \rightarrow 0} {D \over (R/T)^2} .
\end{equation}
The sonic line is subsequently parametrized by the variable 
$Y={\cal M}/RD$ which on the sonic line is just
\begin{equation}
  Y=\frac{-2(\gamma-1)\bSp}{2(\gamma-1) + \gamma\bSp} .
\end{equation}
This approach was also used by Maison \cite{maison} for studying the more
general problem of non-universality in non-self-similar gravitational 
collapse. He  used a set of variables 
$\left(A_M, B_M, \tilde{\rho}_M, v_M\right)$. In their renormalization group 
approach to the critical behaviour of spherically symmetric collapse, Hara 
{\it et al} \cite{hara} used a set of variables 
$\left(N_H, A_H, \omega_H\right)$ 
very similar to the variables of Maison (however, they solve the constraint 
globally). In the self-similar case, these sets of variables are given by
\begin{eqnarray}
  A^2_M&=&A_H^{-1}=\frac{\left(1+\bSp\right)^2-\ba^2}{\bk} , \quad
  B^2_M=N_H^{-2}=\frac{\ba^2}{\left(1+\bSp\right)^2} , \\
  \tilde{\rho}_M&=&\frac{\omega_H}{A_H}= 
  \frac{\left(1-u^2\right)\bSp\ba}{\gamma\bk u} , \quad
  v_M=V_H=v_R . \nonumber
\end{eqnarray}

\end{document}